\documentclass[
aps,
12pt,
final,
notitlepage,
oneside,
onecolumn,
nobibnotes,
nofootinbib,
superscriptaddress,
noshowpacs,
centertags]
{revtex4}
\usepackage{lipsum}
\usepackage{epsfig}
\usepackage{amsmath}
\usepackage{amssymb}
\usepackage{array}
\usepackage{graphicx,url}
\usepackage[colorlinks]{hyperref}
\usepackage[english]{babel}   
\usepackage[utf8]{inputenc}

\begin{document}

\title{Interaction of Domain Walls with Scalar Particles in the Early Universe}

\author{\firstname{D.~P.~}\surname{Filippov}}

\email{filippov.danila.p@gmail.com}
\affiliation{
National Research Nuclear University MEPhI (Moscow Engineering Physics Institute), Moscow, Russia
}
\author{\firstname{A.~A.~}\surname{Kirillov}}
\email{AAKirillov@mephi.ru}
\affiliation{
National Research Nuclear University MEPhI (Moscow Engineering Physics Institute), Moscow, Russia
}

%\date{Received December 13, 2024; revised March 25, 2025; accepted March 25, 2025}

%\today

\begin{abstract}
\textbf{Abstract} -- The formation of solitons (such as closed domain walls) in the super-Early Universe is predicted in a number of theories of the formation of primordial black holes. However, the interaction of particles of the surrounding medium with the solitons should affect their dynamics. In the paper, we consider the interaction between domain walls and scalar particles which can play a role of dark matter. It is shown that when the temperature of the scalar particle gas, caused by the expansion of the Universe, decreases below a certain threshold value, the wall abruptly becomes opaque and locks particles inside itself.
We discuss the dynamics of a single domain wall taking into account pressure of scalar particles locked inside a closed wall.
It is shown, this effect leads to a time delay of domain wall collapse and the deferred formation of primordial black holes.
\end{abstract}

\maketitle

\section{Introduction}

The recent observations made by the Hubble \cite{litlink1} and James Webb \cite{litlink2, litlink3} telescopes confirm the existence of supermassive black holes in the early Universe, whose formation mechanism at high redshifts remains unknown. 
The idea that black holes may have primordial (non-stellar) origin was proposed about six decades ago and has since remained the subject of active study by many scientific groups around the world.
The mechanism of primordial black holes (PBHs) formation as a result of collapse of closed domain walls (DWs) seems to be promising \cite{litlink4, litlink5}.
In the early Universe, domain walls could have formed due to of spontaneous symmetry breaking during the dynamics of a scalar field  having a certain potential \cite{litlink6}. 
Quantum fluctuations of such scalar field during the inflation might lead to the suitable initial conditions for the formation of closed DWs (bubbles) \cite{litlink7, litlink8, litlink9}. 
After the inflation, the horizon $r_h$ evolves as $2t$ while the wall size $r$ expands as $\sqrt{t}$. 
Therefore, at some instant, a domain wall becomes causally connected and begins to contract due to surface tension. 
In the absence of medium interaction with a domain wall, the latter could collapse into a primordial black hole. 
However, the interaction could slow down the collapse of a DW and cause delayed PBH formation due to increasing gas pressure.
Moreover, self-interaction and DW collapse may be a strong source of gravitational waves that can be observed in future gravitational wave experiments \cite{litlink10, litlink11, litlink12, litlink13}. 

The interaction of DWs with surrounding particles has been previously discussed in \cite{litlink14}, where the analytical form of the reflection coefficient for fermion fields was obtained.
The same approach has been applied to study interactions with dark photons \cite{litlink15} and axion-like particles \cite{litlink16}. 
The reflection coefficient of scalar fields was discussed in \cite{litlink17}. 
The friction of domain walls in the surrounding plasma was considered in \cite{litlink5, litlink18}.

In this work, we discuss the dynamics of a single domain wall taking into account pressure of scalar dark matter particles and how this affects the formation of PBHs.

\section{Reflection coefficient}

Let us consider the interaction between the scalar field $\varphi$ which could play a role of the cold dark matter particles (CDM) and DW formed during the scalar field $\phi$ evolution.
Following the model \cite{litlink5}, the domain wall is formed during evolution of the complex scalar field $\phi = r e^{i\theta}$ with the potential 
\begin{equation}
  \label{eq:wall}
  V = \dfrac{1}{4} \bigg( \phi^{*}\phi-\frac{f^2}{2} \bigg)^2 
    + \Lambda^4 ( 1 - \cos\theta )
  .
\end{equation}
Here $r$ is the radial component of the complex field, while $\theta$ is its phase. The parameter $f$ is the value of the field at which the vacuum state occurs, and $\Lambda$ is the small parameter leading to the symmetry breaking.
The last term of \eqref{eq:wall} arising during the quantum renormalization is very small ($\Lambda \ll f \sim H$) and plays a role only after inflation.

At the end of inflation, the field $\phi$ is near the ground state and has the form $\phi = \frac{1}{\sqrt{2}} f e^{i \theta}$. 
After the symmetry breaking at the RD-stage, the domain wall is formed as a result of the field evolution (see details in \cite{litlink5, litlink4}) and has the spatial dependence
\begin{equation}
  \theta (x) 
    = 4 \arctan 
      \left[
        \exp \left( \dfrac{2x}{d} \right)
      \right],
  \label{eq:theta(x)}
\end{equation}
where $d$ is the wall thickness
\begin{equation}
  d = \dfrac{2f}{\Lambda^2}.
  \label{eq:d}
\end{equation}

\begin{figure}[t]
  \centering
  \includegraphics[width=0.55\linewidth]{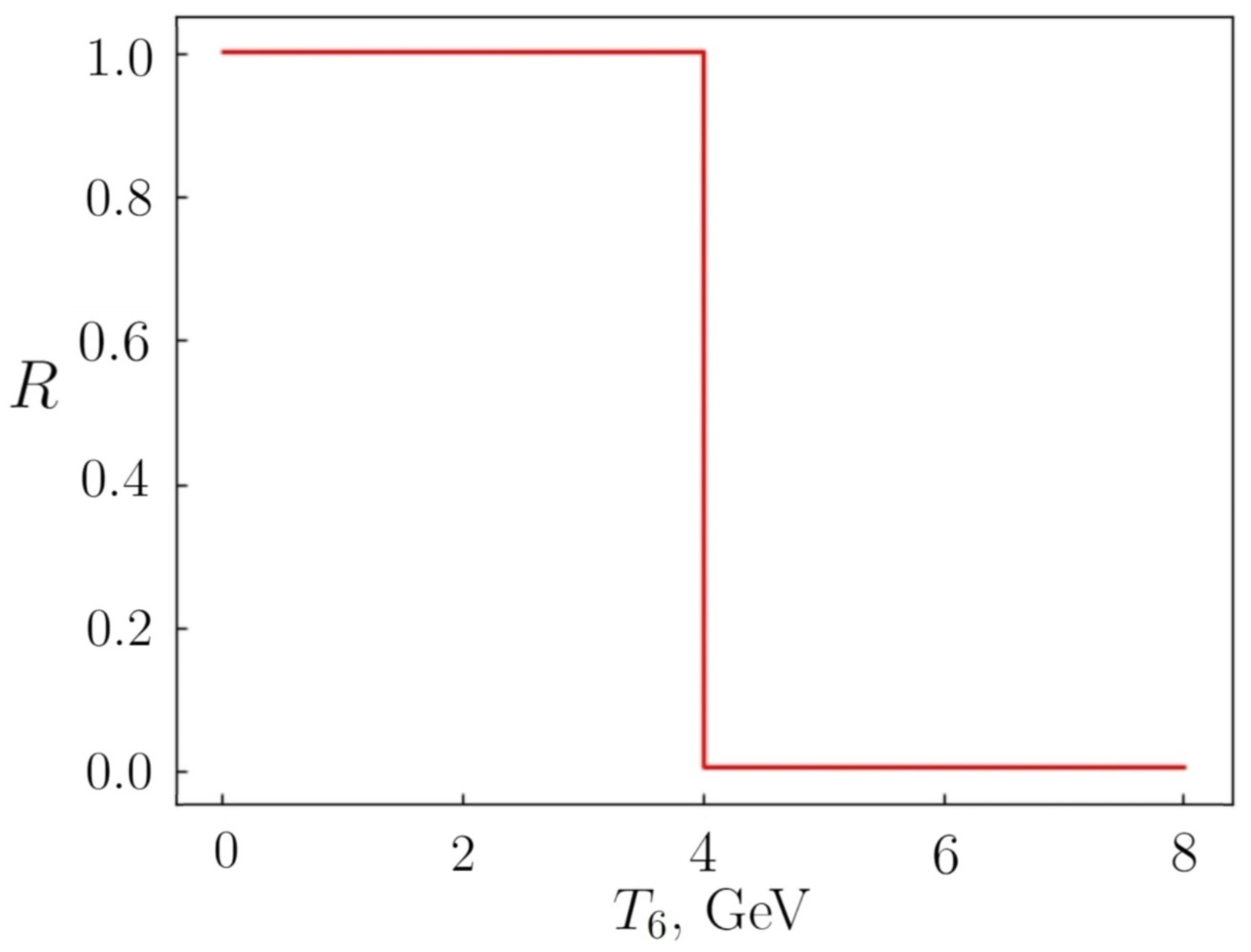}
  \caption{Reflection coefficient $R$ as a function of temperature $T$. $T_6 = 10^6$~GeV.}
  \label{fig.1} 
\end{figure}

Let us choose the interaction term of the Lagrangian with the massive scalar field $\varphi$ as follows
\begin{equation}
    \mathcal{L}_{\text{int}} = \frac{1}{2}\alpha_0(\phi+\phi^{*})\varphi^2 
\end{equation}
and assume the interaction term is small, and the field $\varphi$ does not affect the shape of the DW.
Taking into account Eq.~\eqref{eq:theta(x)}, one could obtain the Klein--Gordorn equation of motion for the scalar field $\varphi$. 
Following the quantum mechanical approach \cite{litlink19} and its realization for scalar fields \cite{litlink12, litlink16, litlink17}, we obtain the reflection coefficient of the scalar field $\varphi$ in the form
\begin{equation}
  \begin{gathered}
    R = \Big[ 1 + \exp\big( q(T) - w \big) \Big]^{-1},
    \\
    q(T) = \pi d \sqrt{p^2(T) + \sqrt{2}\alpha_0 f} , 
    \quad 
    w = \pi \sqrt{2\sqrt{2}\alpha_0 f d^2 -1 } .
    \end{gathered}
  \label{eq:R}
\end{equation}
Here $p$ is the momentum of the scalar particles $\varphi$. 
We assume that the kinetic energy $E_k \approx T$, then the dependence of the momentum on temperature takes the form
\begin{equation}
  p(T) \approx \sqrt{T(t)(T(t)+2m)} ,
\end{equation}
where $m=10^3$~GeV is the mass of the scalar particles. 
The temperature dependence of the reflection coefficient \eqref{eq:R} is shown in Fig.~\ref{fig.1} for the field model with the parameters $f=10^{13}$~GeV, $\Lambda=0.05$~GeV in according to \cite{litlink5} and the interaction constant $\alpha_0 = 1$~GeV. 

One can see, the domain wall abruptly becomes completely opaque for the scalar particles $\varphi$ when the gas temperature $T$ decreases less than the critical value $T_c$:
\begin{equation}
    \label{eq:temperature}
    T < T_c \approx 4\times 10^{6} \text{~GeV} \left(\frac{\alpha_0}{1 \text{~GeV}}\right)^{\frac{1}{2}} \left(\frac{f}{10^{13} \text{~GeV}}\right)^{\frac{1}{2}}.
\end{equation}
Thus, when the Universe cools as a result of the Universe expansion, the scalar particles are locked inside the closed domain wall and must inevitably influence the wall evolution through the gas pressure.

\section{Evolution of the domain wall}

At the end of the inflation, the DW size $r_\text{inf}$ depends on the \textit{e}-fold number $N$ at which the inflationary process generates suitable initial conditions for soliton formation \cite{litlink4, litlink5, litlink7, litlink9} and could be found as
\begin{equation}
    r_\text{inf} = r(t_\text{inf}) = H^{-1} e^{N_\text{inf} - N} ,
\end{equation}
where $N_\text{inf}=60$ is the total number of \textit{e}-folds needed for visible Universe formation, the Hubble parameter at the inflation stage is $H=10^{13}$~GeV, the inflation ends at the moment $t_\text{inf} = N_\text{inf}/H$.

At the radiation stage of the Universe, the cosmological horizon evolves as $2t$, while before entering the cosmological horizon, the domain wall expands as $\sqrt{t}$. Thus, the bubble should cross the cosmological horizon at the instant $t_i$
\begin{equation}
  t_i = \frac{e^{2(N_\text{inf} - N)}}{4 H N_\text{inf}}.
  \label{eq:t_i}
\end{equation}
At this moment, the size of the bubble is equal to the size of the horizon and is estimated to be
\begin{equation}
    r_i = 2 t_i = \dfrac{e^{2(N_\text{inf} - N)}}{2 H N_\text{inf}}. 
    \label{eq:r_i}
\end{equation}

If the gas temperature is less than $T_c$ (Eq.~\eqref{eq:temperature}), the bubble becomes opaque, and after the crossing the cosmological horizon, the dynamics of the domain wall is described by the motion equation \cite{litlink20}
\begin{equation}
  \label{eq:motion}
  \dot{v} (t) = \big( 1-v^2 (t) \big) 
    \bigg( \dfrac{1}{\sigma} \big( P_2(t) - P_1(t) \big) - \dfrac{2\pi}{r(t)} + 3H(t)v(t) \bigg) ,
\end{equation}
where $v$ and $r$  are the speed and radius of the wall, respectively; $\sigma = 4f\Lambda^2$ is the surface energy density of DW, the term $P_2$ is the pressure of scalar particles $\varphi$ locked inside the bubble, while $P_1$ is the pressure of the particles in the surrounding medium (outside the bubble), the term $2\pi/r$ describes the tension of the wall, and the last term is related to the Hubble flow. 
Hereinafter, we assume the radius of the wall is much larger than its thickness $d$ (Eq.~\eqref{eq:d}).

In the thermal equilibrium, a non-relativistic gas of particles can be described by the equation of state $P_i=n_i T_i$, just like the case of a relativistic gas with the accuracy up to a coefficient close to unity.
The change in the number density $n_i$ of the scalar particles $\varphi$ can occur for three reasons. The first one is their annihilation which could be described as
\begin{equation}
  \dot{n}_i(t) = - \dfrac{1}{2} \langle \sigma v \rangle n_i ^2(t),
\end{equation}
where $\langle \sigma v \rangle$ is the velocity averaged annihilation cross-section of scalar particles. 
The second one is related to the change in the volume of the bubble:
\begin{equation}
  \dot{n}_i(t) = - 3 n_i(t) \dfrac{v(t)}{r(t)}.
\end{equation}
The last reason is due to the Hubble flow (outside the bubble):
\begin{equation}
  \dot{n}_i(t) = - 3 n_i(t) H(t) .
\end{equation}

We assume that scalar particles $\varphi$ do not interact with the surrounding matter and there are no reactions increasing their number density. 
%Since particles are locked inside the bubble, their state is adiabatic.
The temperature of the scalar particles inside the closed wall changes only by decreasing or increasing the volume of the bubble and can be found from the first law of thermodynamics in the ideal gas approximation.
The inner energy of the gas is 
\begin{equation}
  \textit{d}{U} 
    \approx \frac{3}{2} \textit{d}(PV)
    \approx \frac{3}{2} nV \textit{d}T ,
\end{equation}
and the gas work takes the form
\begin{equation}
  \delta A 
    =  P \textit{d}V
    =  3nTV \dfrac{\textit{d}r}{r} .
\end{equation}
The first law of thermodynamics for adiabatic process gives
\begin{equation}
  \textit{d}U = - \delta A ,
  \quad \Rightarrow \quad
  % \frac{\df{T}}{T} = - \frac{2}{3} \frac{\df{V}}{V} = -2 \frac{\df{r}}{r(t)} ,
  %   \quad 
  %   \Rightarrow
  %   \quad 
    \dot{T} \approx - 2 T \dfrac{v(t)}{r(t)}.
\end{equation}
The temperature $T_2$ and the number density $n_2$ of the gas inside the bubble have the form
\begin{equation}
  \begin{aligned}
    \dot{T_2}(t)&=  - 2 T_2(t) \cfrac{v(t)}{r(t)},
    \\
    \dot{n_2}(t)&= - \dfrac{1}{2} \langle \sigma v \rangle n_2^2(t) - 3 n_2(t) \cfrac{v(t)}{r(t)} .
  \end{aligned}
  \label{eq:inside}
\end{equation}
The temperature and the number density of the gas outside the bubble decrease due to the expansion of the Universe. 
Then, given that at the RD stage the Hubble parameter evolves as $H = 1/2t$, the dynamic variables of the gas outside the wall are found to be
\begin{equation}
  \begin{aligned}
    \dot{T_1}(t)&=  - \frac{T_1(t)}{t},
    \\
    \dot{n_1}(t)&= - \dfrac{1}{2} \langle \sigma v \rangle n_1^2(t) - \dfrac{3}{2}\dfrac{ n_1(t)}{ t}.
  \end{aligned}
  \label{eq:outside}
\end{equation}

Now, let us consider the initial conditions for Eqs.~\eqref{eq:motion}, \eqref{eq:inside} and \eqref{eq:outside}. 
The initial conditions for Eq.~\eqref{eq:motion} take the form $r(t_i) = r_i$ and $v(t_i) = H(t_i) r_i = 0.9 $, where $t_i$ is chosen as the moment of crossing the cosmological horizon, Eq.~\eqref{eq:t_i}.

The initial number density $n_i$ of CDM particles at the moment $t_i$ is estimated to be
\begin{equation}
    n_i = \frac{\Omega_{\textrm{CDM,0}} \, \rho_{c,0}}{m} (z_i+1)^3 , 
\end{equation}
where $\rho_{c,0} = 4.8\times 10^{-6}$~GeV~cm$^{-3}$ is the present critical density of the Universe, while $\Omega_{\textrm{CDM,0}} = 2.7 \times 10^{-1}$ is the present relative density of CDM \cite{litlink21}. 
The initial redshift $z_i$ could be found from the following
\begin{equation}
 t_i = \int\limits^\infty_{z_i} \frac{\textit{d}z}{(z+1) H(z)}.
\end{equation}
Here Hubble parameter $H(z)$ is
\begin{equation}
    H(z) = H_0 \sqrt{ \Omega_{r,0}(z+1)^4 + \Omega_{m,0} (z+1)^3 + \Omega_{\Lambda,0} }.
\end{equation}
The present day Hubble parameter is $H_0 = 67$~km/s/Mpc, $\Omega_{r,0} = 5.4 \times 10^{-5}$ is the present day radiation density, $\Omega_{m,0} = 3.2 \times 10^{-1}$ is the pressureless matter density, and $\Omega_{\Lambda,0} = 6.9 \times 10^{-1}$ is the dark energy density \cite{litlink21}.

Before CDM decoupling from the primordial plasma, the temperature of DM particles is equal to the temperature of photons.
For the initial values $t_i$, of interest the initial plasma temperature is on the order of $0.1$-$10$~GeV, while the decoupling temperature for CDM particles with the mass $m = 10^3$~GeV is $T_d \approx 2$~MeV \cite{litlink22}. 
Therefore, the initial temperature of the scalar particles could be found as $T_i \approx T_{\gamma, 0} (z_i+1)$, where $T_{\gamma, 0} = 2.7$~K is the present CMB temperature. 

\begin{figure}[tp]
  \centering
  \includegraphics[width=0.6\linewidth]{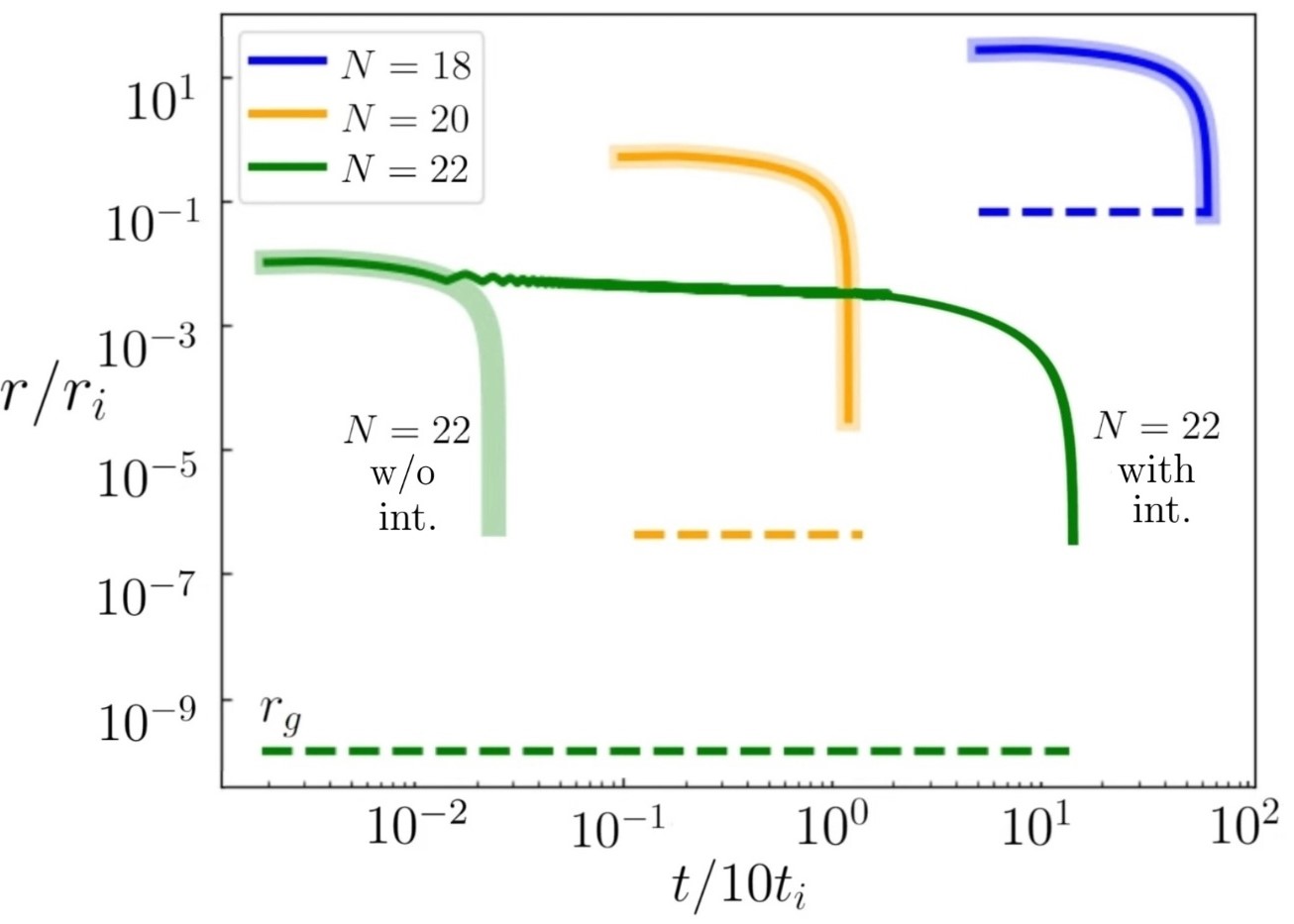}
  \caption{Solid lines represent the cases when the domain walls (formed at $N$-th \textit{e}-fold) interact with the scalar particles, while transparent lines show the cases without interactions. The dashed horizontal lines are the gravitational radii of DWs for each case. The normalization constants are $r_i = 10^6$~cm and $t_i=10^{-5}$~s.}
  \label{fig.2}
\end{figure}

At the moment $t_i$, the CDM temperature $T_1$ and number density $n_1$ inside and $T_2$ and $n_2$ outside the closed DW are equal.
The result of the numerical solution of Eqs.~\eqref{eq:motion}, \eqref{eq:inside} and \eqref{eq:outside} for domain walls formed at $N=18$, $N=20$ and $N = 22$ are shown in Fig.~\ref{fig.2}. 
One can see the oscillations of the radius for the case $N = 22$.
This effect is caused by the balance of two forces: the gas pressure (the term $P_2/\sigma$ in Eq.~\eqref{eq:motion}) and the wall surface tension (the term $2\pi/r$ in Eq.~\eqref{eq:motion}). 

With the selected parameters $f$, $\Lambda$ and $\alpha_0$, the DW formed at $N=18$ collapses to a black hole. 
In the cases $N = 20$ and $N=22$ black holes do not form (since the gravitational radius $r_g < d$), but the particles are heated to the threshold temperature \eqref{eq:temperature} and are able to leave the DW.

The solution of the system of equations for $N=20$ and $N=22$ ends when the domain wall radius $r$ becomes equal to the wall thickness $d$.
To study further dynamics, it is necessary to take into account the self-interaction of the field $\phi$, which is beyond the scope of this paper.

\section{Results and discussion}

Let us consider constraints on the parameters $f$ and $\Lambda$ of the field model \eqref{eq:wall} following from the minimum and maximum mass of a single primordial black hole. 
Note, here we do not discuss the abundance of PBHs in the CDM density which is a separate question and requires consideration of existing observational constraints \cite{litlink23}.

When a DW crosses its gravitational radius $r_g$, a PBH is formed \cite{litlink4, litlink5}.
The restriction on the minimum mass of PBHs is derived from the condition that the gravitational radius is greater than the thickness $d$ of a DW \cite{litlink24}:
\begin{equation}
  \label{eq:Mmin}
  M_\text{min} = 4.8\times10^{-4} M_\odot 
      \left(\frac{f}{10^{13} \, \text{GeV}}\right) 
      \left(\frac{0.05 \, \text{GeV}}{\Lambda}\right)^{2}.
\end{equation}
The maximum possible mass of PBH is determined by the condition that DW does not dominate locally until it enters the cosmological horizon \cite{litlink24}:
\begin{equation}
  \label{eq:Mmax}
    M_\text{max}=7\times10^{8} M_\odot \left(\frac{10^{13} \, \text{GeV}}{f}\right)\left(\frac{0.05 \, \text{GeV}}{\Lambda}\right)^{2}.
\end{equation}
\begin{figure}[t]
  \centering
  \includegraphics[width=0.45\linewidth]{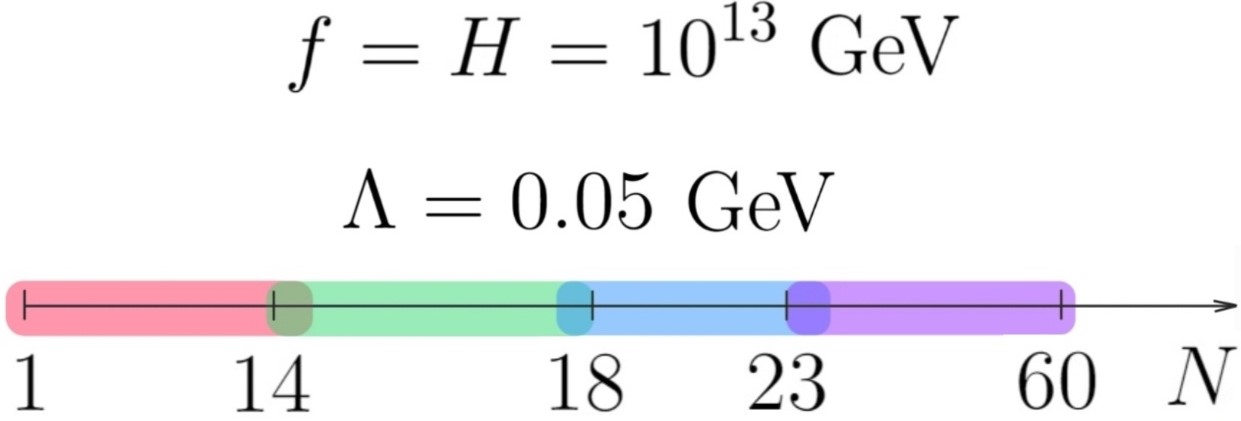}
  \caption{The parameter space of the field model \eqref{eq:wall} vs \textit{e}-fold number $N$ on which a domain wall should begin to form in order to produce PBH (green area). The forbidden region is marked in red ($M > M_\text{max}$, see Eq.~\eqref{eq:Mmax}). A domain wall formed with the parameters marked in the blue area could not produce PBH because $r_g < d$. For the purple area, $r_i < d$.}
  \label{fig.3}
\end{figure}
The allowed region of the potential parameters $f$ and $\Lambda$ depends on the moment of domain wall formation at the inflation stage (\textit{e}-fold number $N$) \cite{litlink4, litlink5}.
Figure~\ref{fig.3} shows at what number $N$ a domain wall should begin to form in order to collapse and produce a black hole in the post-inflationary epoch (green area). 
It is shown that for the production of PBH, the formation of domain walls should start around $N \approx 14$-$18$. 

Thus, it is straightforward to show that there are three bounds for \textit{e}-fold number $N$. 
The first bound follows from the restriction on the upper PBH mass \eqref{eq:Mmax}:
\begin{equation}
\label{eq:N1}
    N > N_1 = \ln{\left(e^{14} \, \frac{\Lambda}{0.05 \, \text{GeV}} \, \sqrt{\frac{f}{10^{13} \, \text{GeV}}}\right)}.
\end{equation}
The second bound follows from the minimum mass of PBH \eqref{eq:Mmin}:
\begin{equation}
\label{eq:N2}
    N < N_2 = \ln{\left(e^{18} \, \frac{\Lambda}{0.05\text{~GeV}}\right)}. 
\end{equation}
The last one follows from the thin-wall approximation which could be interpreted as $r_i \gtrsim 10 d$, where $r_i$ could be found from \eqref{eq:r_i}:
\begin{equation}
\label{eq:N3}
    N < N_3 = \ln{\left(e^{23} \, \frac{\Lambda}{0.05 \, \text{GeV}} \, \sqrt{\frac{10^{13} \, \text{GeV}}{f}}\right)} .
\end{equation}

If DW began to form at $N = 14$-$17$ \textit{e}-folds, its evolution is similar to the case shown on Fig.~\ref{fig.2} for $N = 18$.
The cases $N=19$ and $N=21$ are similar to the case $N=20$.
The case $N=23$ is similar to the case $N=22$ (Fig.~\ref{fig.2}).

\begin{figure}[tp]
  \centering
  \includegraphics[width=0.6\linewidth]{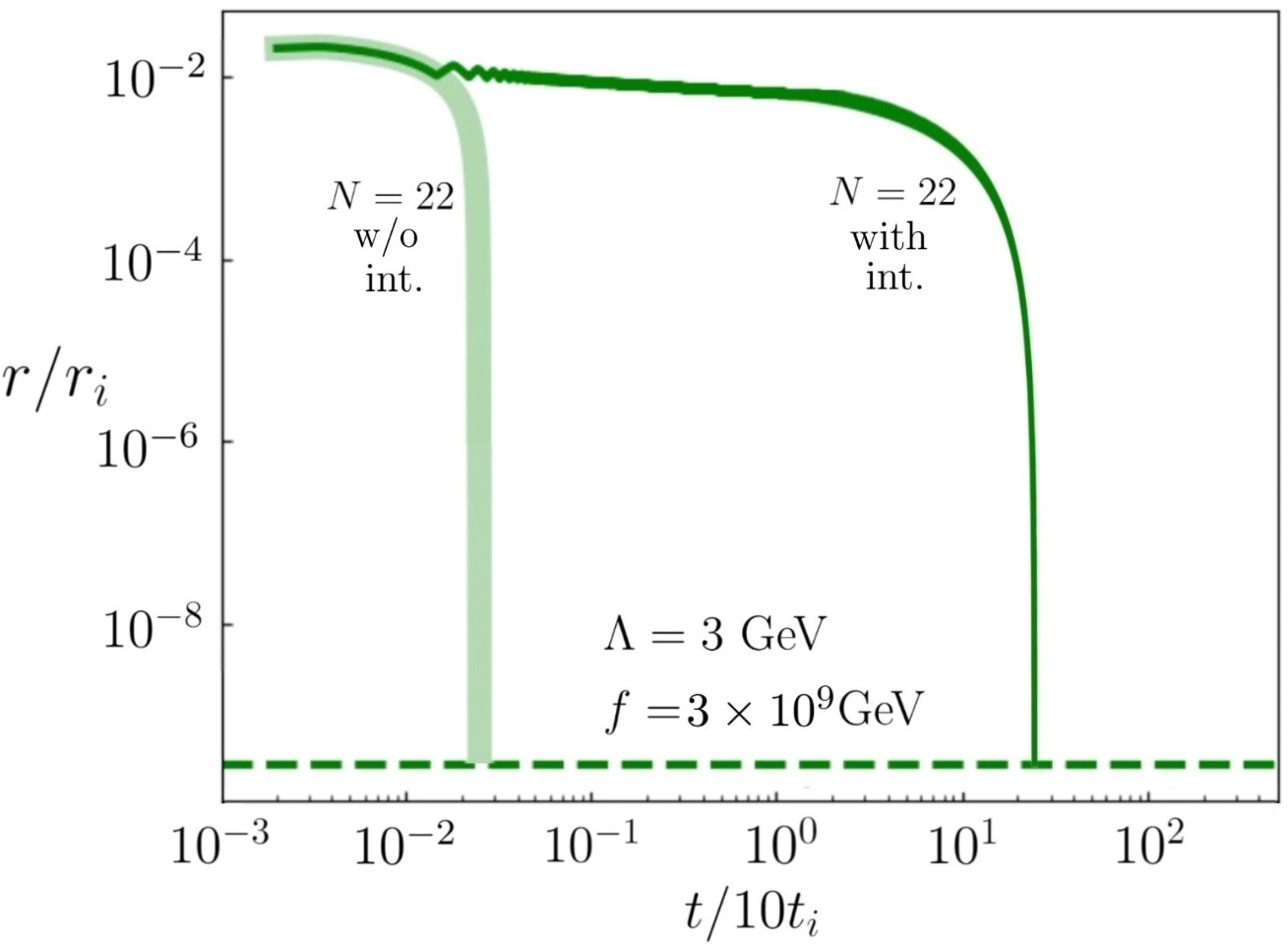}
  \caption{The result of the numerical solution of the system of Eqs.~\eqref{eq:motion}, \eqref{eq:inside} and \eqref{eq:outside} for the case when DW was formed at $N=22$. In contrast to the case presented in Fig.~\ref{fig.2}, the parameters are changed so that DW is able to collapse into a black hole ($f = 3\times10^{9}$~GeV and $\Lambda = 3$~GeV).}
  \label{fig.4} 
\end{figure}

As can be seen from Fig.~\ref{fig.2} and Fig.~\ref{fig.3}, scalar particles do not have a significant effect on PBHs formation with the chosen parameters of the field model \eqref{eq:wall}.
However, if DW began to form at $N=22$-$23$, the pressure of scalar particles locked inside a closed domain wall causes a delay of DW collapse.
The model parameter variation could lead to a time-delayed mechanism of PBHs formation.
For instance, if $\Lambda \ge 3 $~GeV and $f \le 3\times10^9$~GeV (see Eq.~\eqref{eq:N2}), primordial black holes form with the time delay with respect to the case without the gas interaction.
Figure~\ref{fig.4} illustrates the case where the DW generated at $N=22$ forms a PBH with the time delay $\Delta t/10t_i \approx 10^3$.
It is interesting to note that if dark matter particles reach the critical temperature \eqref{eq:temperature} during oscillations that are caused by gas pressure and surface tension of a wall (see Figs.~\ref{fig.2} and \ref{fig.4}), DW collapse is accompanied by emission of dark matter particles, which naturally leads to formation of a DM protohalo around a PBH.
The mechanism of PBHs production due to collapse of closed DWs \cite{litlink4, litlink5} predicts the PBHs formation in the clusters. 
If multiple primordial black holes or their clusters form a single halo, then it becomes possible to detect these objects by their gravitational wave emissions, as suggested in \cite{litlink25}. 
If DM sector has a coupling with Standard Model (SM) particles, the local heating of DM leads to heating of SM particles around DW. 
Such a region can be a source of neutrino radiation, photons and positrons, for which a DW is transparent.

Moreover, if we consider DW interactions with fermions and lock them inside the bubble \cite{litlink14}, we can expect a lot of interesting astrophysical effects.
For example, the bubble could be interpreted as a region with the exotic nucleosynthesis \cite{litlink26} and neutrino cooling \cite{litlink27}. 
If the bubble is long lived, it can be seen as a pseudostar with metal-enriched gas formed by thermonuclear reactions at high temperature.
Very high temperature leads to local recovery of electroweak symmetry breaking and generation of massless SM particles inside the bubble.
The detection of such regions can be an indirect evidence for the existence of domain walls. 

% \clearpage

\begin{acknowledgments}
We are grateful to K.M.Belotsky, V.A.Gani, E.A.Esipova, V.V.Nikulin, S.G.Rubin and V.D.Stasenko for useful discussions and their interest in the work.
The work was funded by the Ministry of Science and Higher Education of the Russian Federation, Project ``New Phenomena in Particle Physics and the Early Universe'' FSWU-2023-0073.
\end{acknowledgments}

\end{document}